\documentclass[preprint,prc,tightenlines,showpacs,superscriptaddress,nofootinbib]{revtex4-1}
\usepackage{amsfonts}
\usepackage{amssymb}
\usepackage{amsmath}
\usepackage{graphicx,color}
\usepackage{dcolumn}
\usepackage{epsfig}
\usepackage{bm}
\usepackage{ulem}

\usepackage{color}

\begin{document}

\title{ Incoherent Pion Production in Neutrino - Deuteron Reactions  }
\author{Jia-Jun Wu}
\affiliation{Physics Division, Argonne National Laboratory, Argonne, Illinois 60439, USA}
\author{T. Sato}
\affiliation{Department of Physics, Osaka University, Toyonaka, Osaka 560-0043, Japan}
\author{T.-S. H. Lee}
\affiliation{Physics Division, Argonne National Laboratory, Argonne, Illinois 60439, USA}

\begin{abstract}
Within the multiple scattering formulation, the incoherent pion production in  neutrino-deuteron reactions
at energies near the $\Delta(1232)$ resonance is investigated.
The calculations include an impulse term and one-loop
contributions from nucleon-nucleon ($NN$) and pion-nucleon ($\pi N$)
 final state interactions.
The input
 amplitudes of $\pi N$ scattering and electroweak pion production reaction
on the nucleon
are generated from
a dynamical model which describes very extensive data of $\pi N$ scattering and
both the electromagnetic and the
weak pion production reactions on the nucleon.
The $NN$ scattering amplitudes are generated from
the Bonn potential.
The validity of the  calculational  procedures is established
 by giving a reasonably good  description of the
data of pion photo-production on the deuteron.
The constructed  model is then applied to predict the
 cross sections of $\nu +  d \rightarrow \mu^- + \pi^+ + n+ p$ and
$\nu +d \rightarrow \mu^-+\pi^0+p+p$ reactions. The importance of  including
the $NN$ final state
interactions  to understand the experimental data of these neutrino-deuteron reactions
is demonstrated. Our results strongly suggest that the spectator approximation
used in the previous
analyses to extract the pion production cross sections on the nucleon from the data on the
deuteron is not valid for the $\nu + d \rightarrow \mu^-+ \pi^+ +n+p$, but is a
good approximation for $\nu + d \rightarrow \mu^-+ \pi^0 + p+p$.
\end{abstract}

\maketitle

\section{introduction}

A precise knowledge of neutrino-nucleus reactions is crucial  in  determining the properties of neutrinos and neutrino interactions, such as the mass hierarchy of neutrinos and CP violation in the lepton sector,
from the data of recent and forthcoming experiments on nuclear targets\cite{form12,alva-14,hyer-k,lbne,lbno,wilkin14}.
In the region of a few GeV neutrino energy where the $\Delta$(1232) resonance plays an important role,
quasi-elastic knock out of nucleons and incoherent single pion production processes
are the main reaction mechanisms of the neutrino-nucleus reactions.
Thus, the starting point of analyzing the neutrino-nucleus
reactions in this energy region is a theoretical model which can describe the
cross sections of the neutrino-induced single pion production on proton ($p$) and neutron ($n$).
These cross sections had been obtained from the experiments on hydrogen and deuterium targets at
Argonne National Laboratory (ANL), and Brookhaven National Laboratory (BNL) and by
European Organization for Nuclear Research (BEBC-CERN)
\cite{cam73,bar79,rad82,kit86,kit90,all80,all90,jones89}.
Various theoretical models\cite{sl-2,sl05,pas04,lal05,lal06,her07,her10,lala10} have been
constructed by fitting these data in recent years.
The uncertainties of these models can bring
systematic errors in the neutrino properties determined from applying these
models to analyze the neutrino-nucleus reaction data.
 Parts of these theoretical uncertainties could also originate from the
about 30\%-40\% differences between the ANL and BNL data,
as discussed\cite{pas04,was06}. However, it seems that this problem has been
resolved\cite{gra09,wilkinson14}.

The cross sections of neutrino-induced single pion production
on the proton target can  be best obtained
from the measurements on the hydrogen target. In practice these cross sections were also
extracted from the analysis
\cite{cam73,bar79,rad82,kit86,kit90,all80,all90} of the combined data from the measurements
on both the hydrogen and the deuterium targets.
The essential assumption of these analyses is that
in the region near the peak of the quasi-free nucleon knock out process,
one of the nucleons in the deuteron does not participate in the reaction mechanism and can
be treated as  a spectator in evaluating the cross sections on the deuteron target.
With the same procedure, the cross sections of the
single pion production  on the
neutron target were also  extracted  from the data on the
deuteron target.
In this work we examine the extent to which this spectator approximation
procedure is valid.

We consider the incoherent single pion production reaction on the deuteron owing to the charged currents:
$\nu + d \rightarrow l^- + \pi^+ + p + n$(CC$1\pi^+$)
and $\nu + d \rightarrow l^- + \pi^0 + p + p$(CC$1\pi^0$).
If the nuclear effects, such as those owing to the nucleon
Fermi motion in the deuteron and the final $\pi NN$ interactions, are neglected,
  the mechanisms of this
reaction can be written as
\begin{eqnarray}
 \nu_\mu + d & \rightarrow &\mu^- + \pi^+ + p + n_s \label{eq:ccpip-1}\\
            & \rightarrow &\mu^- + \pi^+ + n + p_s \label{eq:ccpip-2}\\
            & \rightarrow &\mu^- + \pi^0 + p + p_s  \label{eq:ccpi0}
\end{eqnarray}
where $n_s$ ($p_s$) denotes that the neutron (proton) in the deuteron is assumed to be
the spectator of the reaction processes.
One then expects that
the cross sections for three channels on the $p$ and $n$
can be extracted from the data on the deuteron target.
Thus CC$1\pi^0$ (Eq.(\ref{eq:ccpi0})) will give information
on $\nu + n \rightarrow l^- + p + \pi^0$, while CC$1\pi^+$
(Eqs.(\ref{eq:ccpip-1})-(\ref{eq:ccpip-2})) will give information on
$\nu + p \rightarrow l^- + p + \pi^+$ and  $\nu + n \rightarrow l^- + n + \pi^+$.
However, there is no obvious  reason to justify the neglect of the
$\pi NN$ final state interactions.
It is natural to ask whether the extracted cross sections, in particular the cross sections
on the neutron target, have
the accuracy needed to constraint a model for determining the neutrino properties
from analyzing the data of neutrino-nucleus reactions. The purpose of this paper is to investigate
this important question of current interest. This is also needed to
understand the origins of the difficulties, such as those
reported recently in Refs.\cite{her10,lala10},
in obtaining
a fully consistent theoretical explanation of the cross sections
on both the proton and the neutron targets.

To proceed, we need to start with a model which can describe the
electroweak single pion production
on the nucleon in the $\Delta$ (1232) resonance region.
Among the recent models of neutrino induced pion production reactions
\cite{sl-2,sl05,pas04,lal05,lal06,her07,her10,lala10},
we adopt a dynamical model developed in
Refs.  \cite{sl-2,sl05,sl-1} (called SL model ). This reaction model is
defined by an energy independent Hamiltonian which has vertex interactions describing
 the $\Delta$ (1232) excitation
and non-resonant meson-exchange mechanisms derived from phenomenological Lagrangians
by using\cite{sko,sl-1} a unitary transformation method.
By solving the scattering equations derived from the constructed Hamiltonian, the resulting
reaction amplitudes satisfy unitary condition. The SL model
has been well tested\cite{sl-1} against the data of $\pi N$ scattering and
electromagnetic pion production reactions on the nucleon
in the $\Delta$ (1232) resonance region. It also describes\cite{sl-2} well
the cross sections of neutrino-induced single pion production on $p$ and $n$ from ANL, BNL, and
BEBC-CERN.
The advantage of using the SL model is that we can generate both the electromagnetic
and the neutrino-induced pion production amplitudes within
 the same theoretical framework. Because these two amplitudes contain the same
vector current mechanisms, the application of this model to investigate the
neutrino-induced reactions on nuclear targets, such as the deuteron considered in this work, can be
first tested against the available data of reactions induced by photons and
electrons.

By using the SL model and the  high precision
Bonn nucleon-nucleon potential\cite{bonn} , we
have developed a method for
calculating the cross sections of
 incoherent electroweak pion production on the deuteron within the well-studied
multiple scattering theories\cite{kmt,feshbach,thomas}.
Our calculations include an impulse term and one-loop
contributions from nucleon-nucleon ($NN$) and pion-nucleon ($\pi N$)
final state interactions.
We first establish  our calculation procedures by showing that
the available data  of incoherent pion photo-production reaction on the deuteron can be
described reasonably well.
Our results are fairly consistent with those from the earlier
works\cite{darwish,fix,lev06,sch10,tarasov} on this reaction, as discussed later.
Thus  the developed calculation procedures can be  used reliably
to investigate the $\pi NN$ final state interaction
 effects on the cross sections of neutrino-induced single pion production reactions
on deuteron.

In Sec. II, we recall the formula for calculating the cross sections
of electroweak reactions on hadron targets. Our procedures for calculating the incoherent
pion production amplitudes for the deuteron target are
described in Sec. III.
In Sec. IV we test our approach by investigating
the pion photo-production
reactions on the deuteron.
Our results for the neutrino-induced pion production
reactions on the deuteron are presented in Sec. V.
A summary and discussions are given in Sec. VI.

\section{Formulation for the electroweak reactions on hadrons}

The formula for calculating the cross sections of electroweak reactions on a hadron target
have been well developed in the literature\cite{quiqq}.
For calculations on a nuclear target, it is more convenient to
choose the non-covariant
normalization of states:
$<\vec{p}|\vec{p}^{\,\,'}> = \delta(\vec{p}-\vec{p}^{\,\,'}) $ for  plane wave states,
 and $<\Phi_B|\Phi_B> = 1$ for bound states.
The cross sections of neutrino-induced reactions owing to charged-current (CC)
 can then be written
\begin{eqnarray}
\frac{d\sigma}{d\Omega dE_{l'}} & = &
 \left(\frac{G_F V_{ud}}{\sqrt{2}}\right)^2
\frac{1}{4\pi^2}\frac{|\vec{p}_{l^\prime}|}{ |\vec{p}_l|}
 L^{\mu\nu}W_{\mu\nu}, \label{cc-crs}
\end{eqnarray}
where
$G_F=1.166 \times 10^{-5}GeV^{-2}$ is the Fermi coupling constant
and $V_{ud}=\cos\theta_c$=0.974
 with
$\theta_c$ being the Cabibbo angle.
The lepton tensor $L^{\mu\nu}$ depends only on the
momenta of the initial neutrino ($p_l$) and the final lepton($p_{l'}$)
\begin{eqnarray}
L^{\mu\nu}
 & = & 2[p_l^\mu p_{l'}^\nu + p_l^\nu p_{l'}^\mu
    - g^{\mu\nu}((p_l\cdot p_{l'}) - m_l m_{l'})
    + i \epsilon^{\mu\nu\alpha\beta}p_{l,\alpha}p_{l',\beta} ],  \label{lepton-cc}
\end{eqnarray}
with $\epsilon^{0123}=1$.
The hadron tensor is defined as
\begin{eqnarray}
W^{\mu\nu}&=&\bar{\sum_{i}}\sum_{f}(2\pi)^6\frac{E_T}{M_T}
\delta^4(p_i+q-p_f)
<f|J^\mu(0)|i><f|J^\nu(0)|i>^*
\label{eq:hadron-t}
\end{eqnarray}
where $E_T$ and $M_T$ are the energy and  mass of the target hadron,
$p_i$ and $p_{f}$ are the four-momenta of the initial and
 final states, respectively,
and
$\bar{\sum}_i \sum_f $ the
average over the initial spin of the target and the sum
over the final spins of outgoing particles.

As a comparison, we also write here the formula of
electron scattering cross section:
\begin{eqnarray}
\frac{d\sigma}{d\Omega dE_{l'}} & = &
 (\frac{4\pi\alpha}{Q^2})^2
\frac{1}{4\pi^2}\frac{|\vec{p}_{l^\prime}|}{ |\vec{p}_l|}
 L^{\mu\nu}W_{\mu\nu},
\end{eqnarray}
where $W_{\mu\nu}$ is the same hadron tensor defined in
Eq.(\ref{eq:hadron-t}), and $\alpha=1/137$ is the fine structure constant and $Q^2 = - q^2$ with
$q = p_l - p_{l'}$.
The lepton tensor $L^{\mu\nu}$ is written as
\begin{eqnarray}
L^{\mu\nu}
 & = & \frac{1}{2}[p_l^\mu p_{l'}^\nu + p_l^\nu p_{l'}^\mu
    - g^{\mu\nu}((p_l\cdot p_{l'}) - m_l^2)].
\label{eq:lepton-em}
\end{eqnarray}

It is well known~\cite{quiqq}
that the inclusive differential cross sections of electron and neutrino
induced reactions can be expressed
in terms of structure functions $W_i$.
In the limit of vanishing lepton mass $m_l \sim 0$,
the double differential cross section of inclusive electron scattering
 ($e + d \rightarrow e' + X$) is given as
\begin{eqnarray}
\frac{d^2\sigma}{dE'd\Omega'}
 & = & \frac{4\alpha^2 {E'}^2}{Q^4}[ 2 W_1^{em} \sin^2\theta
+  W_2^{em} \cos^2\theta]. \label{eq:crs-ee-incl}
\end{eqnarray}
Here $\theta$ and $E'$ are the scattering angle and
the energy of the final electron in the target rest frame.
The cross section of the charged current neutrino reaction
($\nu/\bar{\nu} + d \rightarrow l'(\bar{l}') + X$)
is given as
\begin{eqnarray}
\frac{d^2\sigma}{dE'd\Omega'}
 & = & \frac{(G_F V_{ud})^2{E'}^2}
{2\pi^2}[2 W_1^{CC} \sin^2 \theta  + W_2^{CC} \cos^2\theta
\pm W_3^{CC} \frac{\epsilon + \epsilon'}{M_T}\sin^2\theta ].
\label{eq:crs-nu-incl}
\end{eqnarray}
Here the structure functions $W_i$ are defined as
\begin{eqnarray}
W_1^\alpha & = & \frac{1}{2}(W^{\alpha\,11}+W^{\alpha\,22})\label{eq:w1}\\
W_2^\alpha & = & \frac{Q^2}{\vec{q}^2}[W^\alpha_1 + \frac{Q^2}{\vec{q}_c^2} W^{\alpha\,00}]\\
W_3^\alpha & = & - \frac{2M_T}{|\vec{q}|}Im(W^{\alpha\,12}),
\label{eq:struc}
\end{eqnarray}
where $W^{\alpha\,11}$, $W^{\alpha\,22}$, $W^{\alpha\,00}$ and $W^{\alpha\,12}$ are
the components of the
hadron tensor defined by Eq. (\ref{eq:hadron-t})
and
are evaluated by using
the electromagnetic current $J^\mu_{em}$ and
weak charged current $J^\mu=V^{\mu} - A^\mu$
for  $\alpha=em$ and $CC$, respectively.
We use $J^0 + \frac{\omega_C}{Q^2} J\cdot q$
to take into account the non-conservation of the axial vector current
in place of $J^0$ in $W^{00}$ for the neutrino reaction.
The direction of the momentum transfer is chosen to be the $z$ direction:
i.e., $q^\mu = (\omega,0,0,|\vec{q}|)$ in the target rest frame
and $q^\mu = (\omega_C,0,0,|\vec{q}_C|)$ in the center of mass frame
of lepton and target.
One can show that the total cross sections
of the reactions induced by photons can be calculated
only from the transverse parts of hadron tensor defined by Eq.(\ref{eq:w1}):
\begin{eqnarray}
\sigma^{tot} = \frac{4\pi^2\alpha}{E_\gamma}W_1^{em}.
\label{eq:photo-tot}
\end{eqnarray}

The similarity of the cross sections
for the photon, electron and neutrino induced reactions, as seen
 in Eqs. (\ref{eq:crs-ee-incl})-(\ref{eq:photo-tot}),
indicates that one can test the reaction models for the neutrino-induced pion production reactions
by using the data of pion photo- and electro-productions.

Starting with Eqs.(\ref{cc-crs})-(\ref{eq:lepton-em}),
one can also write\cite{deforest,gourdin}
the semi-inclusive cross sections in the forms similar to Eqs.(\ref{eq:crs-ee-incl})
and (\ref{eq:crs-nu-incl}).
For the single pion electro-production reactions on the nucleon,
such a form is well known\cite{sl-2,sl-1}.
 To provide the information which is closely related to the recent experimental
initiatives\cite{wilkin14}, it is more straightforward here to take a numerical approach.
We will use
 directly  the formula
 Eqs.(\ref{cc-crs})-(\ref{eq:hadron-t}) to calculate the exclusive cross sections
of $\nu + d \rightarrow l + \pi + N + N$ and then obtain the semi-inclusive
cross sections by integrating out the appropriate variables of the final $\pi NN$ states.
Our numerical procedure is explained  in the next section.

\section{Calculations for the deuteron target}
Our task is to evaluate the  hadron tensor $W^{\mu\nu}$,
defined by Eq.(\ref{eq:hadron-t}), for $\nu (p_{l}) + d (p_d) \rightarrow l' (p_{l'}) + \pi (k) + N_1 (p_1) + N(p_2)$
in the Laboratory frame in which
the deuteron  with mass $m_d$ is at rest and thus its four-momentum is $p_d=(m_d, \vec{0})$.
Suppressing the  spin and isospin indices, the considered hadron tensor becomes
\begin{eqnarray}
W^{\mu\nu}&=&
 (2\pi)^6\frac{E_d(\vec{p}_d)}{m_d}
\int d\vec{k}\,\, d\vec{p}_1
\delta(E_d(\vec{p}_d) + q^0 - E_\pi(\vec{k}) - E_N(\vec{p}_1)
-E_N(\vec{p}_2))\nonumber \\
&&\times <\Psi^{(-)}_{\vec{k},\vec{p}_1,\vec{p}_2}|J^\mu(0)| \Phi_d>
<\Psi^{(-)}_{\vec{k},\vec{p}_1,\vec{p}_2}|J^\nu(0)|\Phi_d>^*
\label{eq:wmunu}
\end{eqnarray}
where $|\Phi_d>$ is the deuteron bound state, and
$|\Psi^{(-)}_{\vec{k},\vec{p}_1,\vec{p}_2}>$ the $\pi NN$ scattering state.

To proceed, we need to define a model for describing the electromagnetic
and weak reaction mechanisms of reactions on the deuteron.
Such a model can be constructed by
 extending the usual two-nucleon Hamiltonian to include
the Hamiltonian developed in the SL model. It is then straightforward to
apply  the well-established multiple scattering
formulation\cite{kmt,feshbach,thomas} to derive formula for calculating the
the current matrix elements
$<\Psi_{\vec{k},\vec{p}_1,\vec{p}_2}^{(-)}|J^\nu(0)|\Phi_d>$.

Keeping only the terms up to the second order in the multiple scattering expansion,
we have
\begin{eqnarray}
<\Psi_{\vec{k},\vec{p}_1,\vec{p}_2}^{(-)}|J^\nu(0)|\Phi_d>
= <{\vec{k},[\vec{p}_1,\vec{p}_2}]_A|J^{Imp,\nu}(0) + J^{NN,\nu}(0)+J^{\pi N,\nu}(0)
|\Phi_d> \label{eq:t-amp}
\end{eqnarray}
where $|\vec{k},[\vec{p}_1,\vec{p}_2]_A> $ is a $\pi NN$ plane-wave state with an
anti-symmetrized
$NN$ component $[\vec{p}_1,\vec{p}_2]_A$. In
the following subsections, we give expressions for the matrix elements of
the impulse term $J^{Imp,\nu}(0)$, the $NN$ final-state interaction term $J^{NN,\nu}(0)$,
and the $\pi N$ final-state interaction term $J^{\pi N,\nu}(0)$.
For each term,
the corresponding reaction amplitude is the sum of the contributions from
 each nucleon
in the deuteron. We only give the formula to calculate
the contribution from the  nucleon $1$.
The formula can then be used for the full calculations with the
properly anti-symmetrized $NN$ in the deuteron and the final $\pi NN$ states.
This procedure is tedious but straightforward, and thus is not
given in the paper.

\begin{figure}[htbp] \vspace{-0.cm}
\begin{center}
\includegraphics[width=0.5\columnwidth]{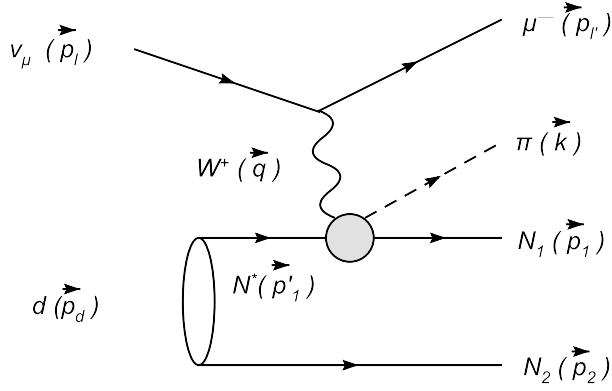}
\caption{Impulse mechanism $J^{Imp,\nu}(0)$ of Eq.(\ref{eq:jcc}).}
 \label{fg:imp}
\end{center}
\end{figure}

\subsection{Impulse term}
With the momenta illustrated in Fig.\ref{fg:imp}, the contribution from
the nucleon $1$ to the impulse term in the deuteron rest frame $p_d=(m_d,\vec{0})$ can be written as
\begin{eqnarray}
<\vec{k},\vec{p}_1,\vec{p}_2|J^{Imp,\nu}(0)|\Phi_{d}> &=&
<\vec{k},\vec{p}_1|j^\nu|\vec{q},\vec{p}_1^{\,\,'}>\times \Phi_d(\vec{p}_c)
\label{eq:jcc}
\end{eqnarray}
where $\vec{p}^{\,\,'}_1=-\vec{p}_2$, and $ \vec{p}_c = (\vec{p_1}^{\,\,'}-\vec{p_2})/{2}$
is the two-nucleon relative momentum, and $j^\nu$ is either
the electromagnetic or weak current associated with the nucleon $1$.
The current matrix element on a single nucleon in Eq.(\ref{eq:jcc}) is calculated\cite{sl-rev} from
\begin{eqnarray}
<\vec{k},\vec{p}_1|j^\nu|\vec{q},\vec{p}^{\,\,'}_1 > &=&
\sqrt{\frac{1}{(2\pi)^{9}}\frac{m^2_N}{2E_\pi(\vec{k})E_N(\vec{p}_1)E_N(\vec{p}_1^{\,\,'})}}
\nonumber \\
&&\times\sum_{\mu}[\Lambda_{lc}(\vec{p_1}+\vec{k},E_N(p_1)+E_\pi(k))]^\nu_{\mu}
<\vec{\kappa}|j^\mu_c(W_c)|\vec{q}_c>
\label{eq:slmx}
\end{eqnarray}
where $\Lambda_{lc}(\vec{p},E)$ is the Lorentz transformation for getting the
current $j^\nu$ in the laboratory frame from
$j^\mu_c$ in the center of mass system of the outgoing
$\pi N$ subsystem.
The vectors $\vec{q}_c$ and $\vec{\kappa}$ denote
 the initial and final three-momentum of $\pi$ in the center of mass system
of the outgoing $\pi N$ subsystem.
Note that
$<\vec{\kappa}|j^\mu_c(W_c)|\vec{q}_c>$ includes the $\pi N$ final state interaction
and thus it depends on the invariant mass $W_c$ of the $\pi N$ subsystem.
By using the three-body approximation  developed in Ref.\cite{thomas},
the invariant mass $W_c$ in Eq.(\ref{eq:slmx}) is calculated from
the energy available to the
$\pi + N$
subsystem.
It is calculated from subtracting the energy $E_N(p_2)$ of the second nucleon
in Fig.\ref{fg:imp} from  the  total energy $\omega + m_d$
of the initial $W^+ + d$  system :
\begin{eqnarray}
W_c = [(\omega+m_d-E_N(p_2))^2-(\vec{p}_1+\vec{k})^{\,\,2}]^{1/2}
\end{eqnarray}
We define the  Lorentz transformation in Eq.(\ref{eq:slmx}) by using the momenta of the
outgoing $\pi N$ subsystem.
Explicitly, we have
\begin{eqnarray}
\Lambda_{lc}(\vec{p},\;E)=\left( \begin{array}{cccc}
\frac{E}{M}           & -\frac{p_{x}}{M}           & -\frac{p_{y}}{M}            & -\frac{p_{z}}{M} \\
-\frac{p_{x}}{M}      & 1+ \frac{p^2_{x}}{M(M+E)}  &  \frac{p_{x}p_{y}}{M(M+E)}  & \frac{p_{x}p_{z}}{M(M+E)} \\
-\frac{p_{y}}{M}      & \frac{p_{y}p_{x}}{M(M+E)}  &  1+\frac{p^2_{y}}{M(M+E)}   & \frac{p_{y}p_{z}}{M(M+E)} \\
-\frac{p_{z}}{M}      & \frac{p_{z}p_{x}}{M(M+E)}  &  \frac{p_{z}p_{y}}{M(M+E)}  & 1+\frac{p^2_{z}}{M(M+E)} \\
\end{array} \right).
\label{eq:lz-lc}
\end{eqnarray}
where $M=[E^2-\vec{p}^{\,\,2}]^{1/2}$.
The  inverse  $[\Lambda_{lc}]^{-1}(\vec{p},E)= \Lambda_{cl}(\vec{p},E)$ is  used to
get the vector $q^\mu_c=(\omega_c, \vec{q}_c)$  from  $q^\mu=(\omega,\vec{q})$, and
$\kappa^\mu=(E_\pi(\vec{\kappa}),\vec{\kappa})$ from $k^\mu=(E_\pi(\vec{k}),\vec{k})$.
We thus have
\begin{eqnarray}
q^\nu_c=\sum_{\mu}[\Lambda_{cl}(\vec{p_1}+\vec{k},E_N(p_1)+E_\pi(k))]^\nu_{\mu}q^\mu \\
\kappa^\nu=\sum_{\mu}[\Lambda_{cl}(\vec{p_1}+\vec{k},E_N(p_1)+E_\pi(k))]^\nu_{\mu}k^\mu
\end{eqnarray}
where
\begin{eqnarray}
\Lambda_{cl}(\vec{p},\;E)=\left( \begin{array}{cccc}
\frac{E}{M}           & \frac{p_{x}}{M}            &  \frac{p_{y}}{M}            & \frac{p_{z}}{M} \\
 \frac{p_{x}}{M}      & 1+ \frac{p^2_{x}}{M(M+E)}  &  \frac{p_{x}p_{y}}{M(M+E)}  & \frac{p_{x}p_{z}}{M(M+E)} \\
 \frac{p_{y}}{M}      & \frac{p_{y}p_{x}}{M(M+E)}  &  1+\frac{p^2_{y}}{M(M+E)}   & \frac{p_{y}p_{z}}{M(M+E)} \\
 \frac{p_{z}}{M}      & \frac{p_{z}p_{x}}{M(M+E)}  &  \frac{p_{z}p_{y}}{M(M+E)}  & 1+\frac{p^2_{z}}{M(M+E)} \\
\end{array} \right).\label{eq:lz-cl}
\end{eqnarray}

As reviewed in Ref.\cite{sl-rev}, we can calculate the current matrix element in the right-hand-side
of Eq.(\ref{eq:slmx}) by the relation
\begin{eqnarray}
-\frac{m_N}{4\pi W_c} <\vec{\kappa}|j^\mu(W_c)|\vec{q}_c>
=\sum_{n} F_n(W_c) O_n(\hat{\kappa},\hat{q}_c,\epsilon^\mu)
\label{eq:mx-cgln}
\end{eqnarray}
where $F_n(W_c)$ is the Chew-Goldberger-Low-Nambu (CGLN)
 amplitudes,  and
$O_n(\hat{\kappa},\hat{q}_c,\epsilon^\mu)$ are the operators in the nucleon spin-space
which can be found in the appendix of Ref.\cite{sl-rev}. We generate the CGLN amplitudes
$F_n(W_c)$ from  the SL model\cite{sl-1,sl-2}.

\begin{figure}[htbp] \vspace{-0.cm}
\begin{center}
\includegraphics[width=0.5\columnwidth]{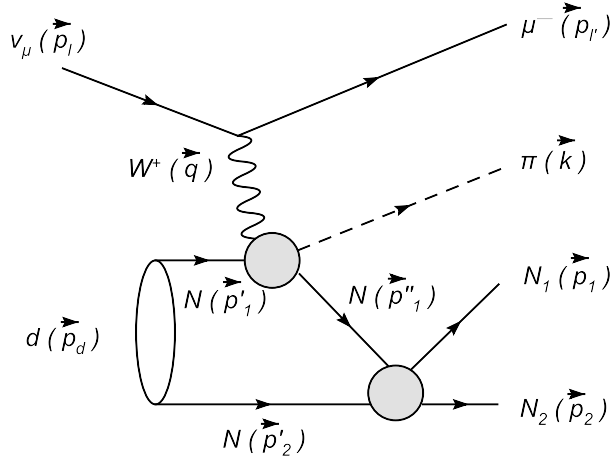}
\caption{The $NN$ final state interaction term $J^{NN,\nu}(0)$ of Eq.(\ref{eq:nn-fsi}).}
 \label{fg:nn-fsi}
\end{center}
\end{figure}

\subsection{$NN$ final-state interaction term}
In the deuteron rest frame, the matrix element of the  $NN$ final-state interaction
term, as illustrated in
Fig.\ref{fg:nn-fsi}, can be written as
\begin{eqnarray}
<\vec{k},\vec{p}_1,\vec{p}_2|J^{NN,\nu}(0)|\Phi_{d}> &=&\int d\vec{p}^{\,\,''}_1
<\vec{p}_1,\vec{p}_2|t_{NN}(E_N(p_1)+E_(p_2))|\vec{p}^{\,\,''}_1,\vec{p}^{\,\,'}_2>
\nonumber \\
&&\times
\frac{1}{E-E_N(p^{''}_1)-E_N(p^{'}_2)-E_\pi(k)+i\epsilon}\nonumber \\
&&\times
<\vec{k},\vec{p}^{\,\,''}_1|j^\nu|\vec{q},\vec{p}_1^{\,\,'}>\,\Phi_d(\vec{p}^{\,\,'}_1)
\label{eq:nn-fsi}
\end{eqnarray}
where $\vec{p}^{\,\,'}_2=-\vec{p}^{\,\,'}_1$, $
\vec{p}^{\,\,'}_1 = \vec{p}^{\,\,''}_1+\vec{k}-\vec{q}$, and
the $NN$ t-matrix is calculated from
\begin{eqnarray}
&&<\vec{p}_1,\vec{p}_2|t_{NN}(E_N(p_1)+E_N(p_2))|\vec{p}^{\,\,''}_1,\vec{p}^{\,\,'}_2>\nonumber\\
&=&\left[\frac{E^2_N(p)E^2_N(p')}{E_N(p_1)E_N(p_2)E_N(p^{''}_1)E_N(p^{'}_2)}\right]^{1/2}
 <\vec{p}|\hat{t}_{NN}(E_c)|\vec{p}^{\,\,'}>
\label{eq:nn-t}
\end{eqnarray}
where $E_c =[ (E_N(p_1)+E_N(p_2))^2-(\vec{p}_1+\vec{p}_2)^2]^{1/2}$ is the energy in the
two-nucleon center of mass system, $\vec{p}$ and $\vec{p}^{\,\,'}$ are the
two-nucleon relative momenta calculated from $(\vec{p}_1, \vec{p}_2)$ and
$(\vec{p}^{\,\,'}_1, \vec{p}^{\,\,'}_2)$, respectively.
The on- and off-shell scattering matrix elements
$<\vec{p}|\hat{t}_{NN}(E_c)|\vec{p}^{\,\,'}>$ in the $NN$ center of mass system
are generated from the Bonn potential.
The  relation Eq.(\ref{eq:nn-t})
 between the two-body matrix elements in the laboratory frame and $NN$ center of
mass frame is commonly used in  multiple scattering calculations\cite{thomas} and
is justified in an investigation of Ref.\cite{bht,lee77}.
The current matrix element $<\vec{k},\vec{p}^{\,\,''}_1|j^\nu|\vec{q},\vec{p}_1^{\,\,'}>$ in
the right-hand side of Eq.(\ref{eq:nn-fsi}) is obtained by
 replacing  $\vec{p}_1$ in Eq.(\ref{eq:slmx}) with $\vec{p}^{\,\,''}_1$.
Here the invariant mass $W_c$ is given as
\begin{eqnarray}
W_c = [(\omega+m_d-E_N(p^{'}_2))^2-(\vec{p}^{\,\,''}_1+\vec{k})^2]^{1/2}.
\end{eqnarray}

\subsection{$\pi N$ final-state interaction term}
\begin{figure}[htbp] \vspace{-0.cm}
\begin{center}
\includegraphics[width=0.5\columnwidth]{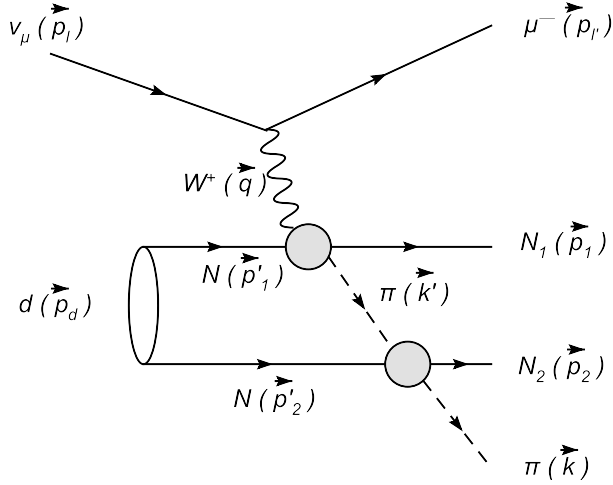}
\caption{The $\pi N$ final state interaction term $J^{\pi N,\nu}(0)$ of Eq.(\ref{eq:pin-fsi}).}
 \label{fg:pin-fsi}
\end{center}
\end{figure}

With the variables given in Fig.\ref{fg:pin-fsi}, the matrix element of
the $\pi N$ final state interaction term in the deuteron rest frame  is
\begin{eqnarray}
<\vec{k},\vec{p}_1,\vec{p}_2|J^{\pi N,\nu}(0)|\Phi_{d}>
 &=&\int d\vec{k}^{\,\,'}
<\vec{k},\vec{p}_2|t_{\pi N}(E_\pi(k)+E_(p_2))|\vec{k}^{\,\,'},\vec{p}^{\,\,'}_2>
\nonumber \\
&&\times
\frac{1}{E-E_N(p_1)-E_N(p^{'}_2)-E_\pi(k')+i\epsilon}\nonumber \\
&&\times
<\vec{k'},\vec{p}_1|j^\nu|\vec{q},\vec{p}_1^{\,\,'}>\,\Phi_d(\vec{p}^{\,\,'}_1)
\label{eq:pin-fsi}
\end{eqnarray}
where $\vec{p}^{\,\,'}_2=-\vec{p}^{\,\,'}_1$, $
\vec{p}^{\,\,'}_1 = \vec{p}_1+\vec{k}^{\,\,'}-\vec{q}$. Similar to the relation Eq.(\ref{eq:nn-t}),
the $\pi N$ t-matrix in Eq.(\ref{eq:pin-fsi}) is calculated from
\begin{eqnarray}
<\vec{k},\vec{p}_2|t_{\pi N}(E_\pi(k)+E_N(p_2)))|\vec{k}^{\,\,'},\vec{p}^{\,\,'}_2>
=[\frac{E_\pi(q_\pi)E_N(q_\pi)E_\pi(q'_\pi)E_N(q'_\pi)}
{E_\pi(k)E_N(p_2)E_N(k')E_N(p^{'}_2)}]^{1/2}
 <\vec{q}_\pi|\hat{t}_{\pi N}(E'_c)|\vec{q}^{\,\,'}_\pi> \nonumber \\
\label{eq:pin-t}
\end{eqnarray}
where $E'_c =[ (E_\pi(k)+E(p_2))^2-(\vec{k}+\vec{p}_2)^2]^{1/2}$ is the energy in the
$\pi N$ center of mass system, $\vec{q}_\pi$ and $\vec{q}^{\,\,'}_\pi$ are the
$\pi N$ relative momenta calculated from $(\vec{k},\vec{p}_2$) and
($\vec{k}^{\,\,'}, \vec{p}^{\,,\,'}_2)$, respectively.
The current matrix element $<\vec{k}^{\,\,'},\vec{p}_1|j^\nu|\vec{q},\vec{p}_1^{\,\,'}>$ in
the right-hand side of Eq.(\ref{eq:pin-fsi}) is obtained by
 replacing  $\vec{k}$ in Eq.(\ref{eq:slmx}) with $\vec{k}^{\,\,'}$.
Here the invariant mass $W_c$ is given as
\begin{eqnarray}
W_c = [(\omega+m_d-E_N(p^{\,\,'}_2))^2
-(\vec{p}_1+\vec{k}^{\,\,'})^2]^{1/2}.
\end{eqnarray}

\section{Test of the model in $\gamma + d \to N+N+\pi $}
To carry out the calculations using the formula described in the previous sections,
we use the SL model
 to generate the current matrix elements $<\kappa|j^\mu|q_c>$ and the
$\pi N$ scattering t-matrix $<q_\pi'|t_{\pi N}|q_\pi>$.
The Bonn potential\cite{bonn} is used to generate
the $NN$ t-matrix $<\vec{p}|t_{NN}|\vec{p}^{\,\,'}>$
 and the deuteron wave function $\phi_d(\vec{p})$.
Thus there is no free parameter in our calculations.
To make realistic predictions of $\nu+ d \rightarrow l + \pi + N +N$
reactions, it is necessary to test our approach by examining the extent to which
the available data of $\gamma + d \rightarrow \pi^-+ p+p\,\,, \pi^0+ n+p$
can be described. Our calculations for this reaction are similar to those
of Refs.\cite{darwish,fix,lev06,sch10,tarasov}, while there are differences between
different approaches
in the formulation and the
input to the calculations.

In addition to the total cross section defined by
Eq.(\ref{eq:photo-tot}),
we  also compare our predictions with the data of  the differential cross sections.
We can derive from Eq.(\ref{eq:photo-tot}) the differential cross sections in
the $\gamma$-d center of mass frame. Including spin and isospin variables explicitly,
we have
\begin{eqnarray}
\frac{d\sigma}{d\Omega_\pi} &=& \int dM_{NN} \frac{4\pi^2\alpha}{2E_\gamma}
(\tilde{W}_{11}+\tilde{W}_{22})
\label{eq:photo-crst}
\end{eqnarray}
where the $\tilde{W}_{\mu\nu}$ can be calculated from $W_{\mu\nu}$ in Eq. (\ref{eq:wmunu}) :
\begin{eqnarray}
\tilde{W}_{\mu\nu}&=&\frac{dW_{\mu\nu}}{d\Omega_{\pi}dM_{NN}}\nonumber\\
&=&
 \frac{(2\pi)^6}{2J_d+1}\frac{E_d(\vec{p}_d)}{m_d}
\int |\vec{k}||\vec{p}^{*}_{NN}|d\Omega^{*}_{NN}\nonumber\\
&&\sum_{M_{J}}\sum_{m_{s_1}}\sum_{m_{s_2}}
\frac{E_N(\vec{p}_1)E_N(\vec{p}_2)E_\pi(\vec{k})}{\omega+m_d-\frac{E_\pi(\vec{k})q}{|\vec{k}|}cos\theta_\pi}
\nonumber\\
&&\times <\vec{k}[\vec{p}_1m_{s_1}m_{\tau_1},\vec{p}_2m_{s_2}m_{\tau_2}]_A|J_\mu(0)| \Phi_d^{J M_{J}, T M_{T}}>\nonumber\\
&&\times <\vec{k}[\vec{p}_1m_{s_1}m_{\tau_1},\vec{p}_2m_{s_2}m_{\tau_2}]_A|J_\nu(0)|\Phi_d^{J M_{J}, T M_{T}}>^*
\label{eq:tildeW}
\end{eqnarray}
where $J^\mu(0)$ is the electromagnetic current.
In the above equation, $\vec{p}^*_{NN}$ and $\Omega^*_{NN}$ are
 the momentum and angle of the nucleon $1$
in the rest frame of
the outgoing $NN$ system.
Here we integrate out the solid angle $\Omega_{NN}^*$;
$\omega$ and $q$ are the energy and three-momentum of the
 momentum-transfer $q^\mu=(\omega,0,0, q)$ to the deuteron;
 $(m_{s_i}, m_{\tau_i})$ are the z-components of the spin and isospin of the $i$-th nucleon,
and ( $J  M_{J}, T M_{T}$)
denote the spin and isospin quantum numbers  of the deuteron.
For the considered photo-production reaction,
we obviously have $\omega=q$. Note that
the $NN$ in the final $\pi NN$ state is anti-symmetrized:
\begin{eqnarray}
|[\vec{p}_1m_{s_1}m_{\tau_1},\vec{p}_2m_{s_2}m_{\tau_2}]_A>
=\frac{1}{\sqrt{2}}\left[|\vec{p}_1m_{s_1}m_{\tau_1},\vec{p}_2m_{s_2}m_{\tau_2}>
- |\vec{p}_2m_{s_2}m_{\tau_2},\vec{p}_1m_{s_1}m_{\tau_1}>\right] \label{eq:ant-nn}
\end{eqnarray}
The deuteron wave function in Eq.(\ref{eq:tildeW})
is (in the deuteron rest frame)
\begin{eqnarray}
|\Phi_d^{J M_{J}, T M_{T}}> &=& \sum_{m_{s_1}m_{\tau_1}} \sum_{m_{s_2}m_{\tau_2}}
\int d \vec{p} |\vec{p} m_{s_1}m_{\tau_1}; -\vec{p} m_{s_2}m_{\tau_2}>
 [\sum_{L=0,2}
<J M_{J}|L\,S\, M_{L}\,M_{S}> \nonumber \\
&&\times <S M_{S}|1/2\,1/2\,m_{s_1}\,m_{s_2}>
<T M_{T}|1/2\,1/2\,m_{\tau_1}\,m_{\tau_2}>Y_{LM_{L}}(\hat{p})]
\end{eqnarray}
 where $S$ ($S=1$)  and $L$ are the spin  and
the orbital angular momentum of two nucleons, respectively.
 We note here that Eqs.(\ref{eq:tildeW}) is independent of lepton kinematical variables except
the momentum-transfer $q=l_p-l_p'=(\omega,\vec{q})$. Thus it can also be used in our
later calculations of $\nu + d \rightarrow l' + \pi + N + N$ by simply using the
weak currents to evaluate the matrix elements of $J^\nu(0)$ in  Eq.(\ref{eq:tildeW}).

\begin{figure}[htbp] \vspace{-0.cm}
\begin{center}
\includegraphics[width=0.6\columnwidth]{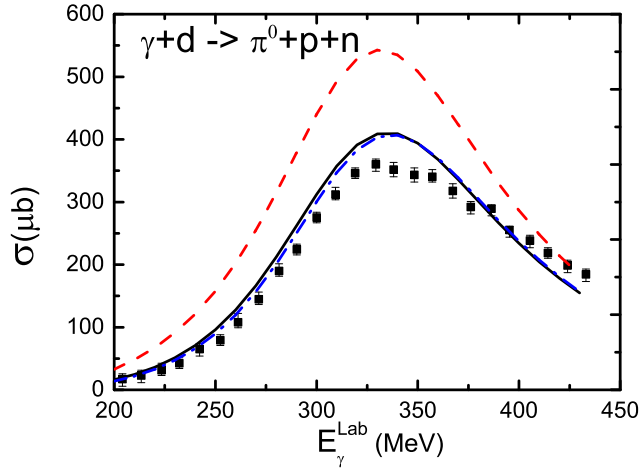}
\caption{(Color online) The total cross sections of $\gamma + d \to \pi^0 + n + p$.
The red dashed, blue dash-dotted, and black solid
curves
represent
only the impulse term , the impulse + ($NN$ final state interaction),
and the impulse + ($NN$ final state interaction) + ($\pi N$ final state interaction), respectively.
Data are from Ref. ~\cite{Krusche:1999tv}.}
 \label{fg:pi0tot}
\end{center}
\end{figure}

\begin{figure}[htbp] \vspace{-0.cm}
\begin{center}
\includegraphics[width=0.6\columnwidth]{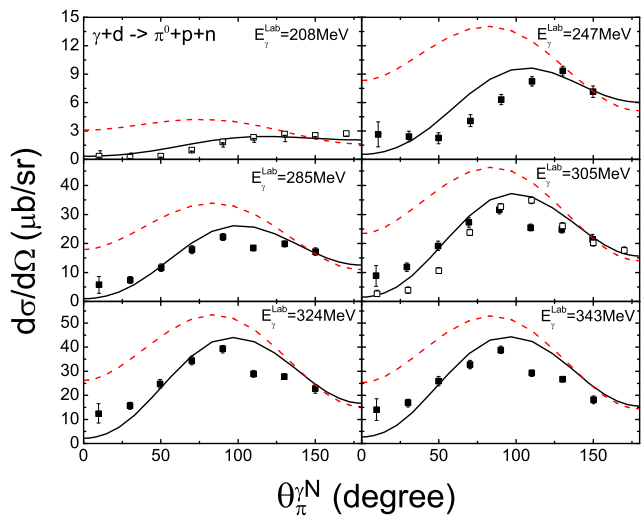}
\caption{(Color online) The calculated differential cross sections (solid curves)
of $\gamma + d \to \pi^0 + n + p$ are compared with the data
from Ref.~\cite{Krusche:1999tv}(solid boxes) and Ref.~\cite{Siodlaczek:2001mh}(open boxes).
The dashed curves are from calculations including only the impulse term $J^{Imp,\nu}(0)$.
Note that\cite{Krusche:1999tv} the experiment data are defined  in the initial $\gamma N$
center of mass system where $N$ is  one of the nucleons which are assumed to be
'frozen' in the deuteron.
This system is equivalent to a system in which
the deuteron momentum $\vec{p}_d$ is related to the photon momentum $\vec{q}$
by $\vec{p}_d = - 2\vec{q}$.}
 \label{fg:pi0dif}
\end{center}
\end{figure}

Our results for the total cross sections of
$\gamma+ d \rightarrow \pi^0+n+ p$ are shown in Fig.\ref{fg:pi0tot}.
 When only the
impulse term $J^{Imp,\nu}$ is included,
we obtain the  dashed curve. It is greatly reduced
to the dot-dashed curve when the $np$ final state interaction term $J^{NN,\nu}$
is added in the calculation. When the $\pi N$ final state interaction
term $J^{\pi N,\nu}$ is also included in our full calculation, we obtain the solid curve.
Clearly, the $np$ re-scattering effects are very large while the $\pi N$
re-scattering give negligible contributions.
In Fig. \ref{fg:pi0dif} we see that the $\pi^0np$ re-scattering effects
bring the  differential
cross sections of $\gamma+ d \rightarrow \pi^0+n+ p$
calculated from keeping only the impulse term (dashed curves) to
  values (solid curves) which are in
reasonable agreement with the data.

\begin{figure}[htbp] \vspace{-0.cm}
\begin{center}
\includegraphics[width=0.6\columnwidth]{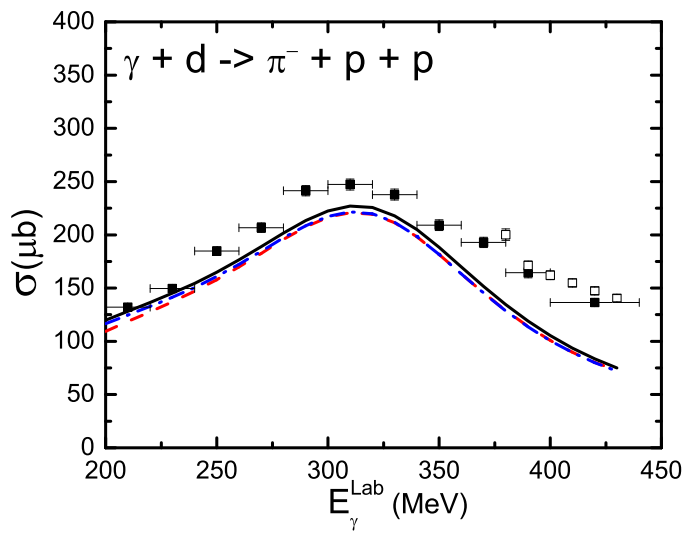}
\caption{(Color online) The total cross sections of $\gamma + d \to \pi^- + p + p$.
The red dashed, blue dash-dotted, and black solid
curves
represent
only the impulse term , the impulse + ($N$ final state interaction),
and the Impulse + ($NN$ final state interaction)
+ ($\pi N$ final state interaction), respectively.
Data are from Ref.~\cite{Benz:1974tt}(Solid boxes) and Ref.~\cite{Asai:1990db}(open boxes).}
 \label{fg:pimtot}
\end{center}
\end{figure}

\begin{figure}[htbp] \vspace{-0.cm}
\begin{center}
\includegraphics[width=0.6\columnwidth]{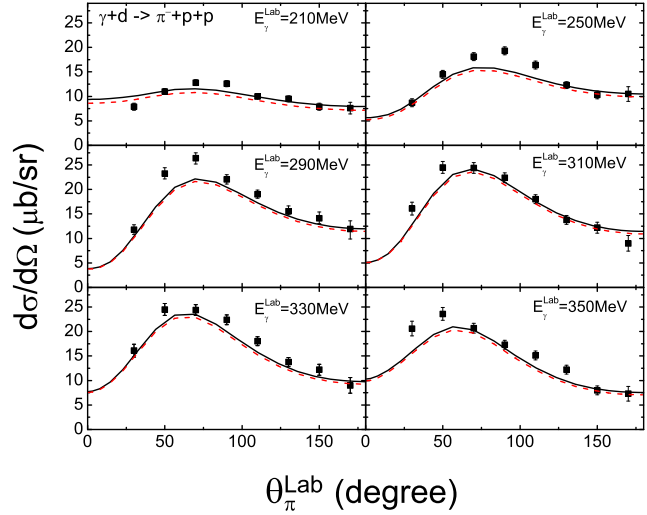}
\caption{(Color online) The  calculated differential cross sections (solid curves)
 of $\gamma + d \to \pi^- + p + p$ in the laboratory frame are compared with the data\cite{Benz:1974tt}.
The dashed curves are from calculations including only the impulse term $J^{Imp,\nu}(0)$. }
 \label{fg:pimdif}
\end{center}
\end{figure}

Similar comparisons for the total cross sections and differential cross sections
for $\gamma+ d \rightarrow \pi^- +p+p$
are shown in Fig.\ref{fg:pimtot} and \ref{fg:pimdif}, respectively.
 Here we see that both the $pp$ and the $\pi N$
final state interactions are weak in this process.
 Comparing these results with those shown in Figs.\ref{fg:pi0tot} and \ref{fg:pi0dif},
we see the
large difference between $np$ and $pp$ final state interactions.
This finding  is consistent with what
was reported
in the previous investigations\cite{darwish,fix,lev06,sch10,tarasov}. It
 perhaps can be understood qualitatively from the properties of
 the initial deuteron wave function
and the final $NN$ wave functions.
We first observe that the final $\pi NN$ interactions are mainly attributable to the s-wave
$NN$ states in the considered energy region.
For $\pi^0np$ final state, the dominant
final $np$ state is $^3S_1+^3D_1$ which has the same quantum number as
the initial deuteron state.
Because the radial wave functions of the deuteron and the scattering state in this
partial wave must be orthogonal to each other,
one expects that the loop integrations over these two wave functions are strongly suppressed
compared with those from the impulse approximation calculations.
In the impulse approximation, the final $np$ state is not orthogonal to the deuteron wave function.
Thus, the large influence of the $np$ re-scattering here is attributable to the elimination of the
spurious coherent contribution in the impulse approximation. A similar discussion can be found
in Ref. \cite{fix}.
However, there is no such orthogonality relation for the $^1S_0$ $pp$ in the $\pi^-pp$.
Consequently the final state interaction effect in
the $\gamma+ d \rightarrow \pi^0+n+p$ is much stronger than that
in the $\gamma +d \rightarrow\pi^-+ p+p$.

We see in
Figs.\ref{fg:pi0tot} - \ref{fg:pimdif}
that  our full calculations (solid curves) are in reasonable
agreement with the data in both the shapes and magnitudes, while some improvements are still needed
in the future.  Thus
our calculation procedure is valid for predicting the $\nu+ d \rightarrow \mu +\pi+ N+ N$
cross sections,  as given in the next section. A more detailed study
of pion photo-production processes is not relevant to our objective here, and therefore
is not further discussed.

\section{Results for $\nu  +d \rightarrow l^- + \pi + N + N$}

Following  the recent experimental initiatives~\cite{wilkin14}, we make predictions for
the incoming muon-neutrino ($\nu_\mu$) energy $E_{\nu_\mu}= 1$ GeV. The
 outgoing muon energy is chosen to be $E_{\mu^-}$ = 550, 600, 650 MeV.
The angle between $\nu_\mu$ and $\mu^-$ is set as
$\theta_{\mu^-}= 25$ degree. This kinematics is chosen to get
maximum values of the predicted cross sections.
The
 coordinate system of the laboratory system
(the rest frame of the deuteron)
is defined as follows: The transfer momentum $\vec{q}$ is
 in the z direction
and the scattering plane of incoming muon-neutrino and outgoing muon is
 the $x-z$ plane.

To proceed, we first calculate the differential cross section
\begin{eqnarray}
\frac{d\sigma}{  dE_{\mu^-} d\Omega_{\mu^-} d\Omega_{\pi} dM_{N\,N}}&=&
 \left(\frac{G_F v_{ud}}{\sqrt{2}}\right)^2 \frac{|\vec{p}_{\mu^-}|}{|\vec{p}_{\nu}|}
\frac{1}{4\pi^2}L^{\mu\nu}
 \tilde{W}_{\mu\nu},
\label{eq:dsigma}
\end{eqnarray}
 where  $\Omega_{\mu^-}$ and $\Omega_{\pi}$ are the solid angles of
outgoing muon and pion, respectively, and $M_{NN}$ is the invariant mass of the outgoing NN system.
The polar and azimuthal angles of pion $\theta_\pi^{Lab}$
and $\phi_\pi$ are the angles from the $z$-axis and the $x$-axis, respectively.
In the following calculation for the
pion angular distribution, we have chosen $\phi_\pi = 0$.
The right-hand-side of Eq.(\ref{eq:dsigma}) can be
calculated by using $L^{\mu\nu}$ of Eq.(\ref{lepton-cc}) and
 $\tilde{W}_{\mu\nu}$ given in Eq.(\ref{eq:tildeW}).
By integrating the $NN$ invariant mass $M_{N\,N}$,
we then obtain  semi-exclusive cross section $\frac{d\sigma}{ dE_{\mu^-} d\Omega_{\mu^-} d\Omega_{\pi}}$.

Our predictions for both
 $\frac{d\sigma}{dE_{\mu^-} d\Omega_{\mu^-} d\Omega_{\pi}}$ and
$\frac{d\sigma}{  dE_{\mu^-} d\Omega_{\mu^-} d\Omega_{\pi} dM_{N\,N}}$ are
presented in the following two subsections.

\subsection{Results of $\nu_\mu + d \to \mu^- + \pi^+ + p + n$}

\begin{figure}[htbp] \vspace{-0.cm}
\begin{center}
\includegraphics[width=0.6\columnwidth]{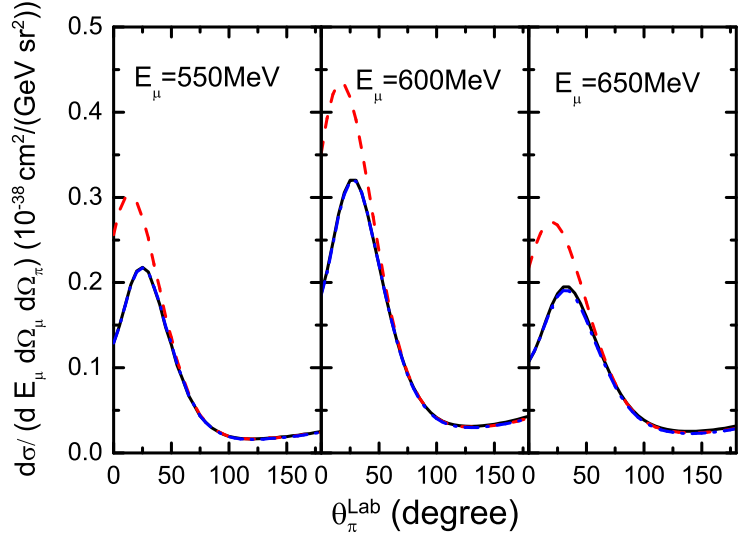}
\caption{(Color online) The
differential cross sections $d\sigma / dE_{\mu^-} d\Omega_{\mu^-} d\Omega_{\pi^+}$ of $\nu_\mu + d \to \mu^- + \pi^+ + p + n$ as function of $\theta_{\pi^+}$ in the laboratory frame
at $E_{\mu^-}$=550, 600, 650 MeV.
The red dashed, blue dash-dotted, and black solid
curves
represent
only the impulse term , the impulse + ($NN$ final state interaction),
and the Impulse + ($NN$ final state interaction) + ($\pi N$ final state interaction), respectively.
The blue dash-dotted and black solid curves are almost indistinguishable because the
$\pi N$ final state interaction effects are very small.
}
 \label{fg:pidif}
\end{center}
\end{figure}

The predicted differential  cross sections
$d\sigma / dE_{\mu^-} d\Omega_{\mu^-} d\Omega_{\pi^+}$
for $\nu_\mu + d \to \mu^- + \pi^+ + p + n$ with $E_\mu = 550, 600, 650$ MeV
 are shown in Fig.\ref{fg:pidif}.
The red dashed curves are from the calculations including only
the impulse term ($J^{Imp, \nu}(0)$) in Eq.(\ref{eq:t-amp}).
When the $NN$ final state interaction term ($J^{NN,\nu}(0)$) is included, the cross sections
are changed to the dot-dashed blue curves. Clearly, the $np$ final state interactions are
significant, in particular in the forward pion angles.
When the $\pi N$ final state interaction
term ($J^{\pi N,\nu}(0)$) term is also included, we obtain our full results denoted as
solid black curves. The small differences between the dot-dashed and solid curves
indicate that the $\pi N$ final state interaction effects are negligible in this
chosen kinematics. This result is similar to what we have observed in our results for
$\gamma + d \rightarrow \pi^0+ n+ p$. This is not surprising because both have the same $np$ scattering
mechanisms.
From the solid black curves in Fig.\ref{fg:pidif},
we see that the cross section with the outgoing muon energy
 $E_{\mu^-} = 600 MeV$ is the largest
in the considered kinematics.

\begin{figure}[htbp] \vspace{-0.cm}
\begin{center}
\includegraphics[width=1.0\columnwidth]{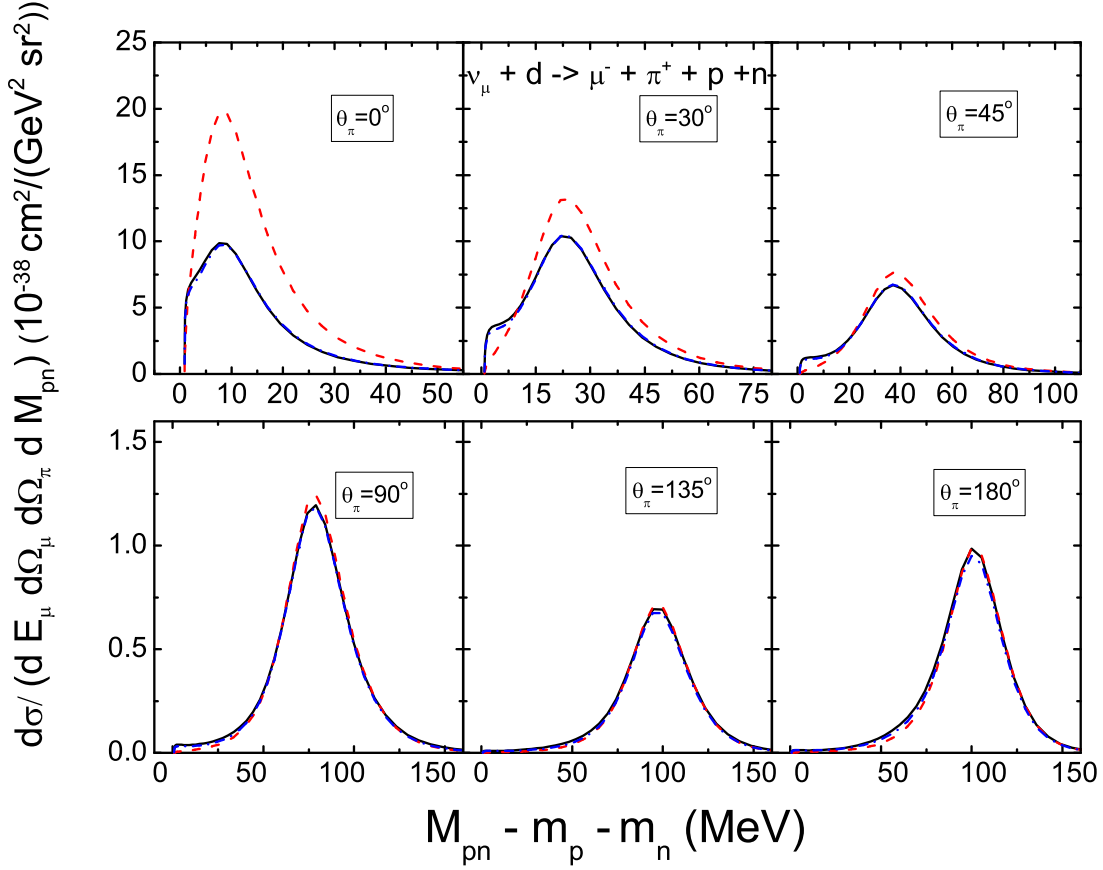}
\caption{(Color online) The differential cross sections $d\sigma / dE_{\mu^-} d\Omega_{\mu^-} d\Omega_{\pi^+} dM_{p\,n}$ of $\nu_\mu + d \to \mu^- + \pi^+ + p + n$ as function of
 $M_{p\,n}$ at several outgoing pion angles $\theta_\pi$. The outgoing muon energy is
 $E_{\mu^-}$=600 MeV.
The red dashed, blue dash-dotted, and black solid curves
represent
only the impulse term , the impulse + ($NN$ final state interaction),
and the Impulse + ($NN$ final state interaction) + ($\pi N$ final state interaction), respectively.
The blue dash-dotted and black solid curves are almost indistinguishable because the
$\pi N$ final state interaction effects are very small.
}
 \label{fg:dif600}
\end{center}
\end{figure}

To understand the angle-dependence of the $np$ final state interaction in Fig.\ref{fg:pidif},
we show
the predicted $NN$ invariant mass distributions
 $d\sigma / dE_{\mu^-} d\Omega_{\mu^-} d\Omega_{\pi^+} dM_{p\,n}$
at $E_\mu=600$ MeV in  Fig.\ref{fg:dif600} for several outgoing pion
angle $\theta_\pi$.
For the forward angles $\theta_\pi \leq 45^0 $,  the $NN$
invariant masses are near the threshold region where the $np$ cross sections are very large
and hence the effects owing to $np$ final state interactions are large.
Furthermore, we find that the shoulders near the threshold
are mainly attributable to
the strong attractive interaction in the  $^3S_1 + ^3D_1$ partial wave of the $pn$ subsystem.
At larger angles $\theta > 90^0$, the allowed
 $NN$ invariant masses are shifted to
the higher mass region around $100$ MeV where the $np$ cross sections are much smaller
 and hence the corresponding $np$ final state interaction effects
are much weaker.

\subsection{Results of $\nu_\mu + d \to \mu^- + \pi^0 + p + p$}

\begin{figure}[htbp] \vspace{-0.cm}
\begin{center}
\includegraphics[width=0.6\columnwidth]{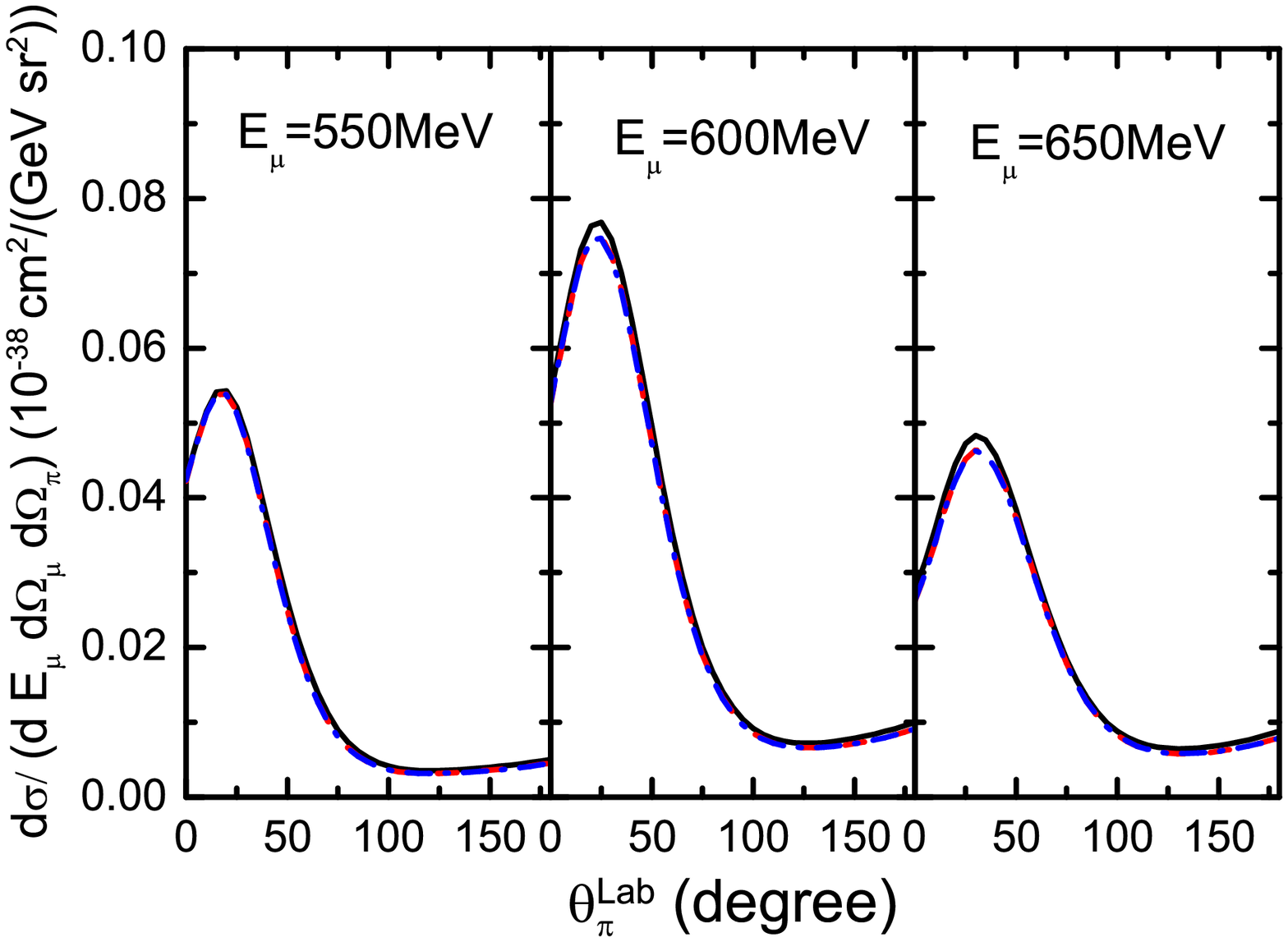}
\caption{(Color online) The differential cross sections $d\sigma / dE_{\mu^-} d\Omega_{\mu^-} d\Omega_{\pi^0}$ of $\nu_\mu + d \to \mu^- + \pi^0 + p + p$ as function of  $\theta_{\pi^0}$ in the laboratory system
 at $E_{\mu^-}$=550, 600, 650 MeV.
The red dashed, blue dash-dotted, and black solid curves
represent
 only the impulse term , the impulse + ($NN$ final state interaction),and the Impulse + ($NN$ final state interaction) + ($\pi N$ final state interaction), respectively.
The blue dash-dotted and black solid curves are almost indistinguishable because the
$\pi N$ final state interaction effects are very small.
}
 \label{fg:pidifpi0}
\end{center}
\end{figure}

\begin{figure}[htbp] \vspace{-0.cm}
\begin{center}
\includegraphics[width=1.0\columnwidth]{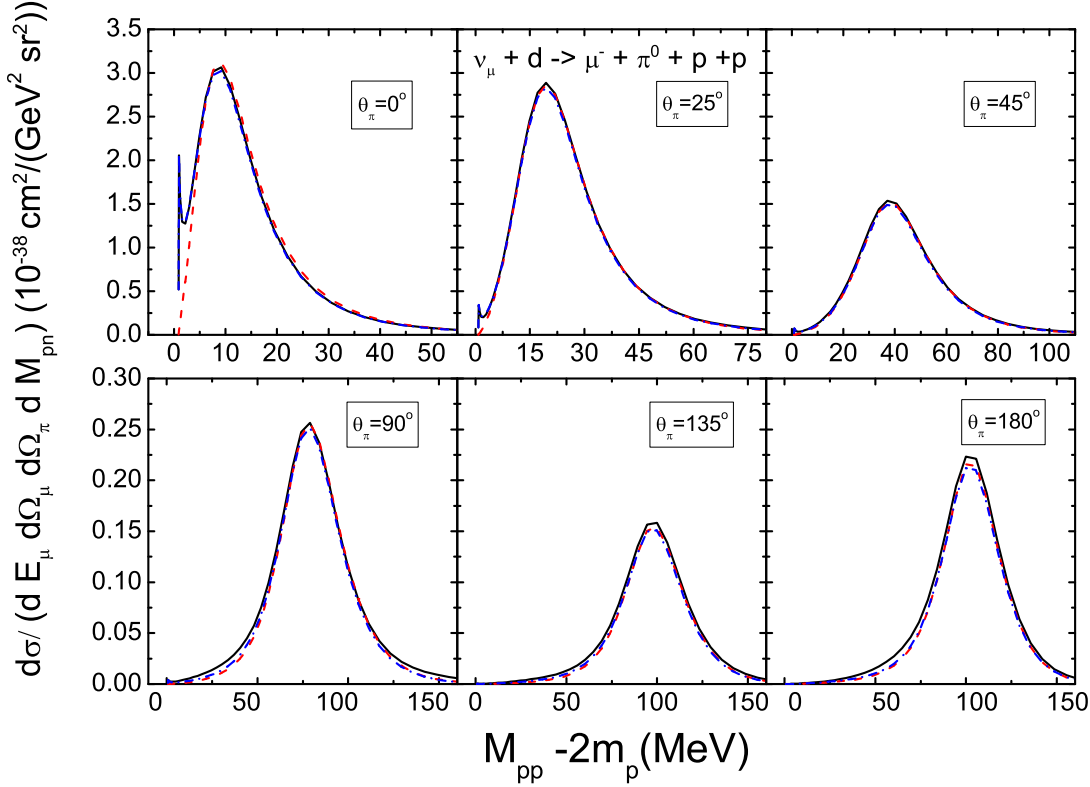}
\caption{(Color online) The differential cross sections $d\sigma / dE_{\mu^-} d\Omega_{\mu^-} d\Omega_{\pi^0} dM_{p\,p}$ of $\nu_\mu + d \to \mu^- + \pi^0 + p + p$ as function of
 $M_{p\,p}$ in the laboratory system. The outgoing muon energy is $E_{\mu^-}$=600 MeV.
The red dashed, blue dash-dotted, and black solid curves
represent
only the impulse term , the impulse + ($NN$ final state interaction),and the Impulse + ($NN$ final state interaction) + ($\pi N$ final state interaction), respectively.
The blue dash-dotted and black solid curves are almost indistinguishable because the
$\pi N$ final state interaction effects are very small.
}
 \label{fg:dif600pi0}
\end{center}
\end{figure}

\begin{figure}[htbp] \vspace{-0.cm}
\begin{center}
\includegraphics[width=0.7\columnwidth]{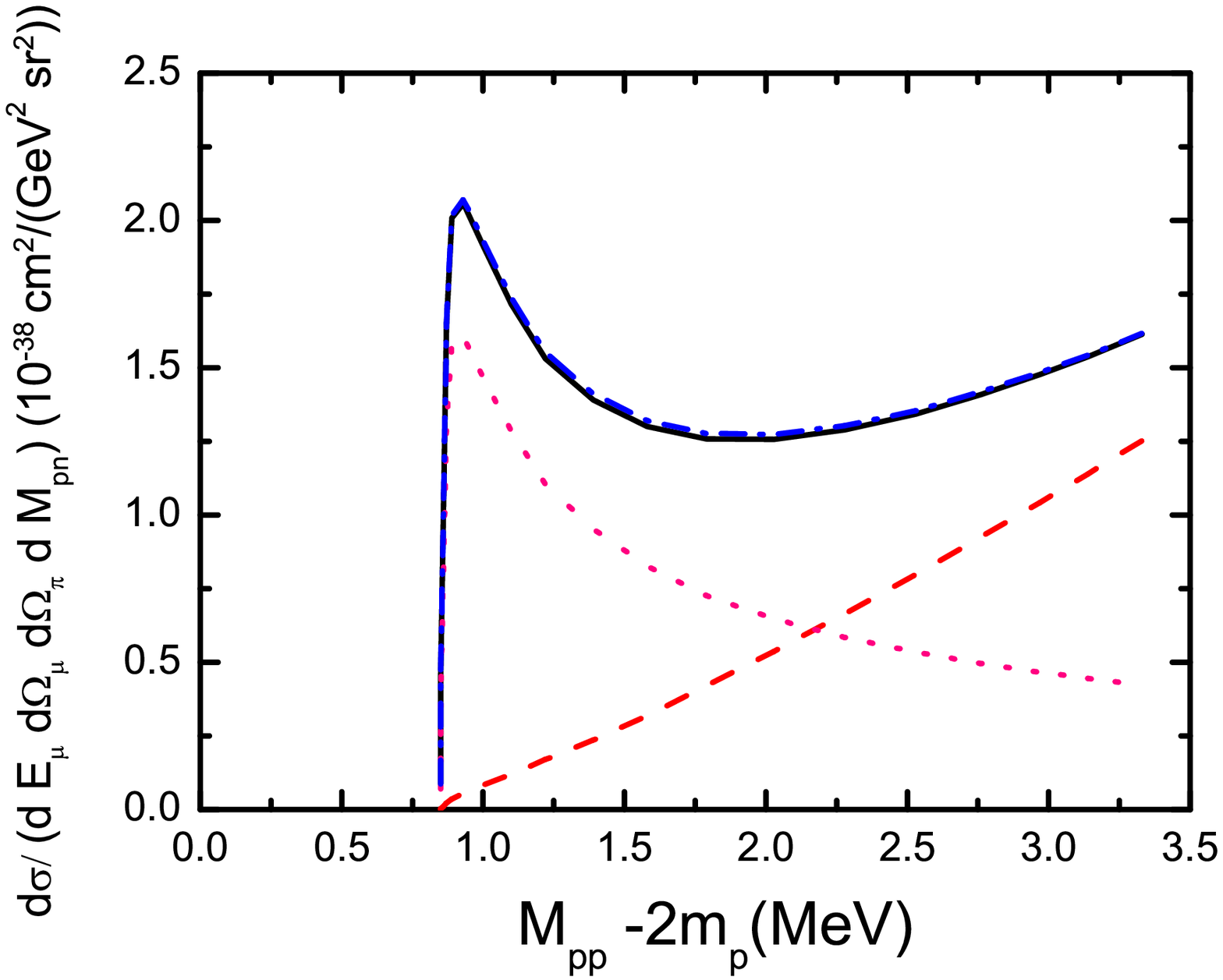}
\caption{The differential cross sections $d\sigma / dE_{\mu^-} d\Omega_{\mu^-} d\Omega_{\pi^0} dM_{p\,p}$ of $\nu_\mu + d \to \mu^- + \pi^0 + p + p$ as function of
 $M_{p\,p}$ in the laboratory system. The outgoing pion angle is $\theta_\pi=0^0$,
and the outgoing muon energy is $E_{\mu^-}$=600 MeV. The red dashed and pink dotted curves are from
calculations including only the Impulse term and only the $NN$ final state interaction term, respectively.
The blue dash-dotted and black solid curves
represent
 the impulse + ($NN$ final state interaction),and the Impulse + ($NN$ final state interaction) + ($\pi N$ final state interaction), respectively.
The blue dash-dotted and black solid curves are almost indistinguishable because the
$\pi N$ final state interaction effects are very small.
}
 \label{fg:dif600pi0thre}
\end{center}
\end{figure}

In Fig.\ref{fg:pidifpi0}, we present the predicted  differential  cross sections of
$d\sigma / dE_{\mu^-} d\Omega_{\mu^-} d\Omega_{\pi^0}$ of the
$\nu_\mu + d \to \mu^- + \pi^0 + p + p$ reaction.
In contrast with the
$\nu_\mu + d \rightarrow \mu^- +\pi^+ + n +p$,
we see that the results (red dashed curves)
from the  calculations  including
only the impulse term are  close to the results (blue dot-dashed curves) including also the
$pp$ final state interaction. The situation here is similar to
what we have observed  in the preview section that the  final state interaction effects
from $np$ scattering are much larger than that from $pp$ scattering.
Comparing the dot-dashed curves and the
black solid  curves from our full calculations, we see that
the $\pi N$ final state interaction effects are negligible. This is also similar
to what we have seen in Fig.\ref{fg:pimtot} for
the $\gamma + d \rightarrow \pi^- + p+p$ process.
The weak $pp$ and $\pi N$ final state interaction effects can also be seen clearly
in Fig.\ref{fg:dif600pi0} for
the $NN$ invariant mass distribution
$d\sigma / dE_{\mu^-} d\Omega_{\mu^-} d\Omega_{\pi^0} dM_{p\,p}$.
The only exception is that
a pronounced sharp peak at the forward
pion angles  $\theta_{\pi^0} = 0^0,\,25^0$.
The origin of this peak can be seen in  Fig.\ref{fg:dif600pi0thre}.
We see that the impulse term (red dashed curve) raises
smoothly from the threshold, while
 the $pp$ final interaction, which is dominated by the
$^1S_0$ partial wave in this very low energy region, generates a peak (pink dotted curve).
A similar discussion has been given in Ref.\cite{fix}.
The Coulomb interaction between two protons is not taken into account in the
present work. Whether this peak will be modified  needs to be investigated in the future.

We note that the cross sections of $\nu +d \rightarrow \mu^- +\pi^0+ p+p$ cross section
is smaller by
 a factor of about 4 than the $\nu + d \rightarrow \mu^- +\pi^+ +n+p$ presented in the previous
subsection.
This is mainly attributable to the fact that
the $\pi^+ pn $ production cross sections include
$\nu_\mu + p \to \mu^- + \pi^+ + p$ and $\nu_\mu + n \to \mu^- + \pi^+ + n$,
while the $\pi^0 np $ production  include only $\nu_\mu + n \to \mu^- + \pi^0 + p$.
Furthermore, the cross section of $\nu_\mu + p \to \mu^- + \pi^+ + p$ is much
larger than that of other two reactions~\cite{sl-2}.

\subsection{Extraction of nucleon cross sections from the deuteron data.}
In Sec. I, we describe
 a procedure that was used in the
previous analyses\cite{cam73,bar79,rad82,kit86,kit90,all80,all90}
to extract the neutrino-induced single pion production
cross sections on the proton and neutron from the data on the deuteron target.
It is based on the assumption that
in the region near the quasi-free peaks, one of the nucleons in the deuteron
is simply a spectator of the reaction mechanisms.
Here we  use our model to examine the extent to which this
procedure is valid.

\begin{figure}[htbp] \vspace{-0.cm}
\begin{center}
\includegraphics[width=0.49\columnwidth]{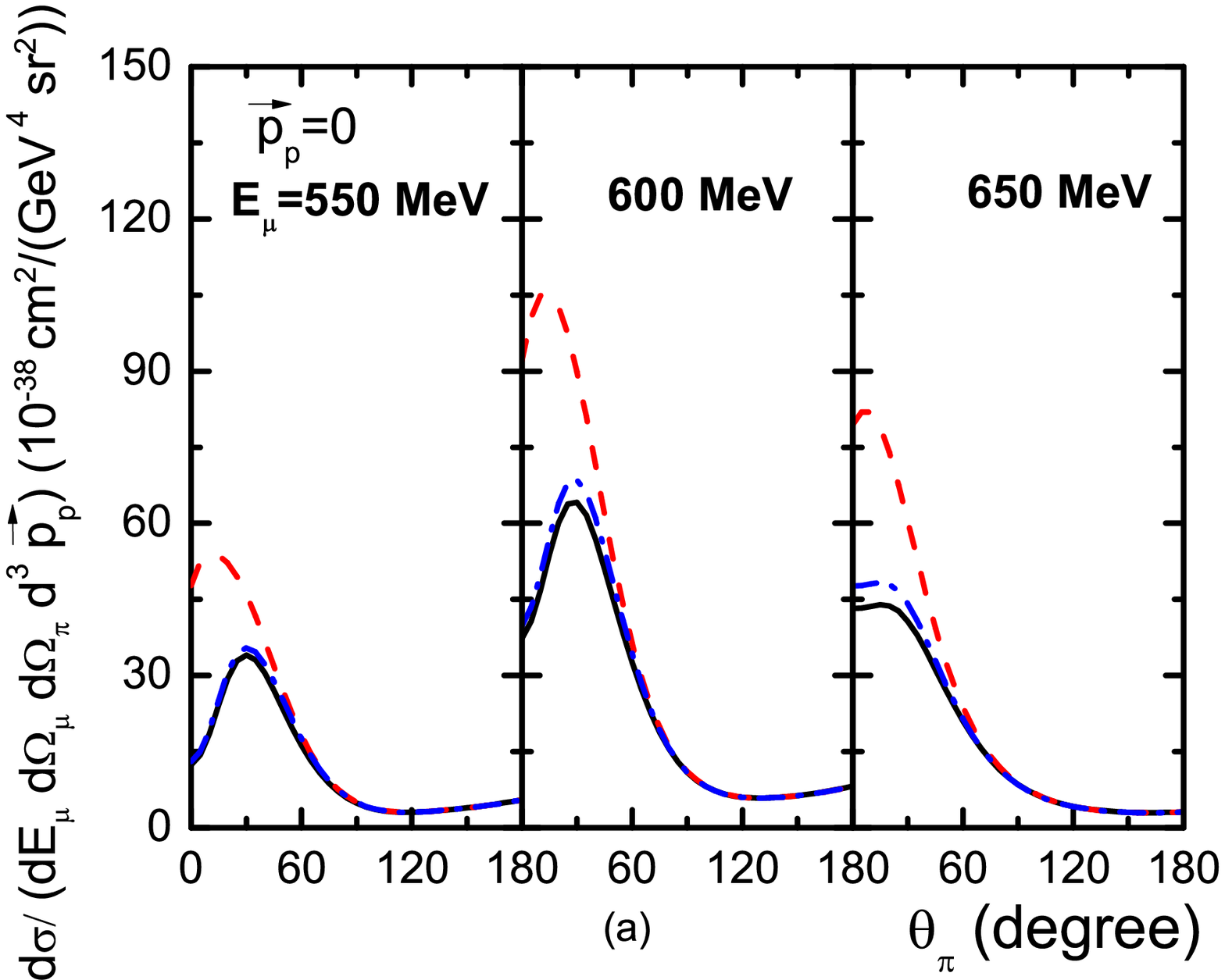}
\includegraphics[width=0.49\columnwidth]{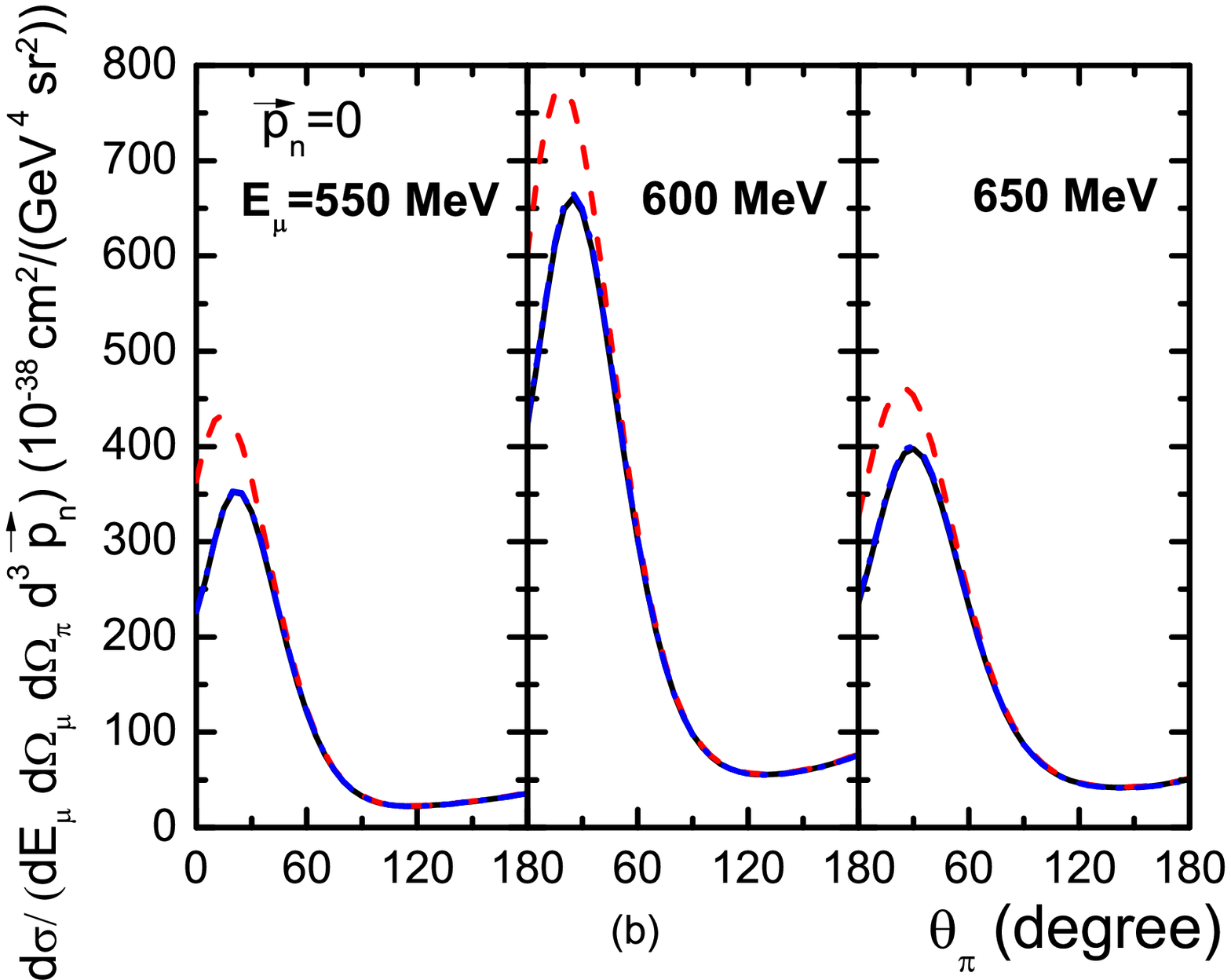}
\caption{(Color online) The differential cross sections
$d\sigma / dE_{\mu^-} d\Omega_{\mu^-} d\Omega_{\pi^+} d^3\vec{p}_N$
of $\nu_\mu + d \to \mu^- + \pi^+ + p + n$
as function of pion scattering angle in the laboratory system.
Panels (a) and (b) figures are for the
proton $\vec{p}_p=0$ and neutron $\vec{p}_n=0$ spectator kinematics, respectively.
The outgoing muon energy is $E_{\mu^-}$=550, 600, and 650MeV.
The red dashed, blue dash-dotted, and black solid  curves are from
calculations including only the Impulse term, Impulse +($NN$ final state interaction),
and Impulse +($NN$ final state interaction)+($\pi N$ final state interaction), respectively.
The blue dash-dotted and black solid curves are almost indistinguishable because the
$\pi N$ final state interaction effects are very small.
}
 \label{fg:pnzero}
\end{center}
\end{figure}

To be specific, we consider the case that
the spectator nucleon is at rest. If there are no final state interactions,
the  $\nu + d \rightarrow l^- +\pi^+ + n+p$ cross section
is only from the pion production on the other nucleon  which is also at rest in the
deuteron rest frame. Then the cross sections measured at the kinematics where the final proton (neutron)
is at rest $\vec{p}_p=0$ ($\vec{p}_n=0$) are simply the cross sections of
$\nu_\mu + n \to \mu^- + \pi^+ + n$ ($\nu_\mu + p \to \mu^- + \pi^+ + p$ ).
The cross sections for this special kinematics can be calculated
  from keeping only the impulse term $J^{imp,\nu}(0)$ in Eq.(\ref{eq:wmunu}).
These
 are the dashed curves in Fig. \ref{fg:pnzero}.
Here we note that the  dashed curves  of $\vec{p}_n=0$ (right)
are almost one order of magnitude larger
than those of  $\vec{p}_p=0$ (left).
This can be understood from the relation
$<\pi^+ p| J_{CC}^\mu|p> = 3 <\pi^+ n| J_{CC}^\mu|n>$
of the charged current contributions
in the isospin $I=3/2$ channel which dominates the reaction cross sections in
 the $\Delta$ (1232) resonance region.

When the $NN$ final-state interaction terms are included, we obtain the dot-dashed curves in
Fig. \ref{fg:pnzero}. The solid curves are obtained  when
the $\pi N$ final state interaction is also
included in the calculations.
Clearly, the $NN$ re-scattering can significantly change the cross sections while the
$\pi N$ re-scattering  effects are weak.
It is also important to note that
the $NN$ re-scattering effects on the cross sections for $\vec{p}_p=0$
are rather different than for $\vec{p}_n=0$.

The results shown in Fig. \ref{fg:pnzero} strongly suggest that the
spectator assumption used in
the previous analyses\cite{cam73,bar79,rad82,kit86,kit90,all80,all90} is
not valid for the CC$1\pi^+$ process $\nu + d \rightarrow \mu^-+ \pi^++ n+p$.
This result is attributable to
the large $np$ re-scattering effects, as explained in Sec. IV and V.A.

We have also examined the results for $p_s=0$ for the
CC$1\pi^0$ process $\nu + d \rightarrow \mu^-+ \pi^0+ p+p$.
Here we find that the spectator assumption is a good approximation for
extracting the cross section on the nucleons from the deuteron target.
This is, of course, attributable to the weak $pp$ final state interactions,
as can be seen in Fig.\ref{fg:pidifpi0}.

\section{Summary and discussions}
We have developed an approach to predict the cross sections of
electroweak pion production on the deuteron  in the energy region near the
$\Delta(1232)$ resonance.
 Within the multiple scattering formulation\cite{kmt, feshbach},
the calculations include the impulse term and the
one-loop contributions from $NN$ and $\pi N$ final state interactions.
 The current matrix elements on the nucleon and the $\pi N$ scattering
amplitudes are generated from
the SL model of electroweak  pion production on the nucleon
developed in  Refs.\cite{sl-1,sl-2}. The $NN$ scattering amplitudes and the deuteron bound state
wave function are generated from Bonn potential\cite{bonn}.There is no free parameter
in the calculations.

We first test the validity of the constructed model by investigating the
pion
photo-production on the nucleon. The predicted cross sections are in reasonable
agreement with the available data, while some further improvements are needed.
The importance of the $np$ final-state interactions is demonstrated, in agreement with
the results of Ref.\cite{darwish,fix,lev06,sch10,tarasov}.

To provide information for the recent experimental initiatives\cite{wilkin14},
we make predictions for the incoming mu on-neutrino energy $E_{\nu}= 1$ GeV.
The differential cross sections for the
 outgoing muon energies $E_{\mu^-}$ = 550, 600, 650 MeV and lepton scattering
angle $\theta_{\nu_\mu,\mu}=25^0$ are presented. It is found that the $np$ final state
interaction effects are very large in determining the differential cross sections
of
 $\nu_\mu+ d \rightarrow \mu^- + \pi^+ + n+ p$ in the region where
the outgoing pions are in the forward angles with respect to the incoming neutrinos.
However, the $pp$ final state
interaction effect is found to be  weak in
 the $\nu_\mu+ d \rightarrow \mu^- +\pi^0 +p+p$ except that
it generates a sharp peak at energies very near the $pp$ threshold.
The $\pi N$ final state interactions are found to be weak in both processes.

Our results strongly suggest that
  the spectator approximation procedure used in the previous
analyses to extract the pion production cross sections on the nucleon from the data on the
deuteron is not valid for the $\nu + d \rightarrow \mu^-+ \pi^+ + +n+p$, but is a
good approximation for $\nu + d \rightarrow \mu^-+ \pi^0 + +p+p$.

In the present calculations, we have not included the contributions from the
exclusive
$\nu_\mu + d \rightarrow \mu^- + \pi^+ + d$ processes. Furthermore, only
the loop contributions from $NN$ and $\pi N$ final state interactions are included.
To improve the accuracy of our predictions for analyzing future experiments on neutrino
properties,
it is necessary to make further developments of the model constructed in this work.
It will be highly desirable to perform calculations by
extending the  unitary $\pi NN$ reaction models, as reviewed in \cite{pinn-rev}, to
include the  electroweak currents. Specifically, this can be done by extending
the unitary $\pi NN$ calculations of Ref.\cite{pinn-lm}
to include the electroweak currents of the  SL model\cite{sl-1,sl-2}.
It will be also
important to apply our approach to investigate neutrino-deuteron reactions
in the higher energy region where the higher mass nucleon
resonances play important roles.
Such  an investigation can be performed when the coupled-channel
model of $\pi N$ and $\gamma N$ reactions developed in Ref.\cite{knls13}
has been extended to
include weak axial currents\cite{ns14}.
Our effort in these directions will be reported elsewhere.

\clearpage
\begin{acknowledgments}
We thank B. Krusche for his help in explaining the
data of Ref.\cite{Krusche:1999tv}.
This work was supported by the U.S. Department of Energy, Office of Nuclear Physics Division,
under Contract No. DE-AC02-06CH11357, and
JSPS KAKENHI Grant Nos. 24540273 and 25105010.
This research used resources of the National Energy Research Scientific Computing Center,
which is supported by the Office of Science of the U.S. Department of Energy
under Contract No. DE-AC02-05CH11231, and resources provided on the Blues and/or Fusion,
high-performance computing cluster operated by the Laboratory Computing Resource Center
at Argonne National Laboratory.
\end{acknowledgments}

\end{document}